\documentclass[onecolumn,secnumarabic,amssymb, nobibnotes, aps,notitlepage, pr,superscriptaddress,10pt]{revtex4-2}

\usepackage{amsmath,amssymb}
\usepackage{stackengine,graphicx}
    \graphicspath{{images/}} 
\usepackage{natbib}
\usepackage[dvipsnames]{xcolor}
\usepackage{bbm}
\usepackage{booktabs}

\usepackage{graphicx}
\usepackage{enumitem}
\usepackage{makecell}
\usepackage{verbatim}

\usepackage{xurl}

\PassOptionsToPackage{hyphens}{url}\usepackage{hyperref}
\usepackage{soul}

\usepackage{bbm}

\usepackage{hyperref}
\hypersetup{
    colorlinks=true,
    linkcolor=blue,
    filecolor=magenta,      
    urlcolor=cyan,
    citecolor=black
}

\def\l{\left}
\def\r{\right}

\usepackage{array}
\newcolumntype{L}[1]{>{\raggedright\let\newline\\\arraybackslash\hspace{0pt}}m{#1}}
\newcolumntype{C}[1]{>{\centering\let\newline\\\arraybackslash\hspace{0pt}}m{#1}}
\newcolumntype{R}[1]{>{\raggedleft\let\newline\\\arraybackslash\hspace{0pt}}m{#1}}

\usepackage{accents}

\newcommand{\lucoh}{Department of Community and Population health, College of Health, Lehigh University, Bethlehem, Pennsylvania, United States of America}
\newcommand{\lucas}{Department of Mathematics, College of Arts and Science, Lehigh University, Bethlehem, Pennsylvania, United States of America}

\date{\today} 

\usepackage[paperwidth=8.5in,paperheight=11.0in,
  left=1.in,right=1.in,top=1.5in,bottom=1.5in,
  includefoot,heightrounded]{geometry}

\begin{document}

\title{A Cluster-Aggregate-Pool (CAP) Ensemble Algorithm for Improved Forecast Performance of influenza-like illness}

\author{Ningxi~Wei}
\affiliation{\lucas}

\author{Xinze~Zhou}
\affiliation{\lucas}

\author{Wei-Min~Huang}
\affiliation{\lucas}

\author{Thomas~McAndrew$^{*}$}
\affiliation{\lucoh}
\email{mcandrew@lehigh.edu}

\setlength{\parskip}{0.5em}
\setlength{\parindent}{0pt}

\begin{abstract}
Seasonal influenza causes on average  425,000 hospitalizations and 32,000 deaths per year in the United States.
Forecasts of influenza-like illness (ILI)---a surrogate for the proportion of patients infected with influenza---support public health decision making.
The goal of an ensemble forecast of ILI is to increase accuracy and calibration compared to individual forecasts and to provide a single, cohesive prediction of future influenza.
However, an ensemble may be composed of models that produce similar forecasts, causing issues with ensemble forecast performance and non-identifiability.  
To improve upon the above issues we propose a novel Cluster-Aggregate-Pool or `CAP' ensemble algorithm that first clusters together individual forecasts, aggregates individual models that belong to the same cluster into a single forecast~(called a cluster forecast), and then pools together cluster forecasts via a linear pool.
When compared to a non-CAP approach, we find that a CAP ensemble improves calibration by approximately 10\% while maintaining similar accuracy to non-CAP alternatives. 
In addition, our CAP algorithm (i) generalizes past ensemble work associated with influenza forecasting and introduces a framework for future ensemble work, (ii) automatically accounts for missing forecasts from individual models, (iii) allows public health officials to participate in the ensemble by assigning individual models to clusters, and (iv) provide an additional signal about when peak influenza may be near.  
\end{abstract}

\maketitle

\clearpage
\section{Introduction}

Seasonal influenza~(Flu) has caused in the United States 9m-41m illnesses, 140k-710k hospitalizations, and 12k-52k deaths annually between 2010 to 2020~\cite{fluBurden,rolfes2018annual}.
Flu disproportionately impacts children under the age of 3 and adults over the age of 65~\cite{10.1542/peds.2013-1493}.
Influenza burden in the United States is estimated to cost \$87B annually in total costs~(\$10B in direct medical costs)~\cite{molinari2007annual}.
This estimate accounts for costs due to hospitalization, deaths, productivity loss as a result of symptoms, and decreased economic activity~\cite{putri2018economic,molinari2007annual}.

Influenza-like illness~(ILI)---a fever plus sore throat/cough---is a syndrome that correlates with laboratory confirmed influenza and can be monitored to help guide public health response~\cite{reich2019collaborative,lutz2019applying}. 
A patient admitted to a healthcare facility is diagnosed with ILI if he/she present with a fever above 37.8 degrees celsius and a cough or sore throat~\cite{iliDef}. 
ILI is a syndromic classification and may include other infectious agents that present symptoms similar to influenza such as SARS-CoV-2 and Respiratory syncytial virus~(RSV)~\cite{iliDef}.
Public health officials monitor ILI to determine proper allocation of resources to hospitals, timely advice for when the public should be vaccinated, and, in severe cases, whether to take actions such as quarantine~\cite{qualls2017community}.

Accurate forecasts of ILI complement surveillance efforts by providing advanced warning of potential changes in burden due to influenza~\cite{reich2019accuracy,mcgowan2019collaborative,osthus2022fast}.
The importance of forecasting the trajectory of infectious agents is highlighted by the FluSight challenge hosted by the Centers for Disease Control and Prevention~(CDC)~\cite{reich2019accuracy,mcgowan2019collaborative}.
The FluSight challenge encourages the development of innovative forecasting models to predict the spread and impact of seasonal influenza, and promotes collaboration among experts in epidemiology, biostatistics, and public health to anticipate and prepare for flu outbreaks.~\cite{lutz2019applying, biggerstaff2021improving}.

Models to forecast ILI can be separated into individual (or component) models and multi-model ensembles. 

Component models train on past observations of percent ILI (number of cases of ILI divided by number of recorded hospital visits) and potentially external data  to generate predictive densities over future percent ILI~\cite{reich2019accuracy,mcgowan2019collaborative,doi:10.1073/pnas.1208772109,SINGH202231,10.1371/journal.pcbi.1004382}. 
Examples of external datasources used in training include commuter patterns, vaccine data, and viral strain data~\cite{pei2020aggregating,wang2019defsi,10.1371/journal.pcbi.1004382}.  
Component models can be phenomenological, taking advantage of statistical correlations between ILI and additional covariates; or mechanistic, supposing a deterministic relationship for how a set of observed and latent variables evolve over time~\cite{pei2020aggregating,wang2019defsi,10.1371/journal.pcbi.1004382,doi:10.1073/pnas.1208772109,SINGH202231}.

A multi-model ensemble takes as input: (i) a set of component model forecasts, (ii) past/present observed ILI, and generates as output a single forecast~\cite{yamana2017individual,ray2018prediction,reich2019accuracy,ray2020ensemble,mcandrew2021adaptively}. 
Past work has shown multi-model ensemble forecasts perform well in practice~\cite{yamana2017individual}.
A multi-model ensemble assumes no individual model can perfectly capture infectious disease dynamics.
Instead, a multi-model ensemble attempts to incorporate many distinct possibilities for how an infectious agent may evolve over time~\cite{ray2018prediction,mcandrew2021adaptively}. 
An ensemble provides a single message for public health officials to interpret~\cite{reich2019accuracy,mcgowan2019collaborative,ray2020ensemble}.
Weaknesses associated with multi-model ensembles are (i) they often assume that forecasts from component models are statistically independent, (ii) these models usually assume that all component models will continue to generate forecasts for the entire season (i.e. no missing forecasts), and (iii) that past performance of component models is indicative of future performance.

We propose a novel framework to ensemble modeling called \textbf{Cluster-Aggregate-Pool~(CAP)}. 
CAP partitions component model forecasts into sets~(Cluster), aggregates each set of forecasts into a single, representative forecast called a cluster forecast~(Aggregate), and then combines cluster forecasts into a single, ensemble forecast~(Pool).  
Our CAP approach to ensemble modeling may better satisfy assumptions of independence, shows similar or improved forecast performance to a non-CAP ensemble, and is able to account for missing component model forecasts in real-time.

\section{Epidemiological and forecast data}
\subsection{Weighted influenza-like illness (ILI)}
Percent influenza-like illness~(ILI) is collected weekly~$(t)$ at the state level~$s$ during an influenza season~(epidemic week 40 of year $Y$ to epidemic week 20 of year $Y+1$)~\cite{reich2019collaborative,mcgowan2019collaborative}. 
ILI is defined as the number of patients who were diagnosed with ILI at a healthcare facility that is a part of U.S.~Outpatient ILI Surveillance Network~(ILINet) divided by the total number of patients who visited a healthcare facility for any reason~\cite{iliDef}.

Weighted percent ILI~(wILI) is computed for each of the 10 Health and Human Service regions~(HHS) and at the US national level and is computed as
\begin{align}
    \text{wILI}_{r} = \sum_{s \in S_{r}} w_{s} \cdot \text{ILI}_{s}
\end{align}
where $S_{r}$ is the set of states that belong to HHS region $r$, $w_{s}$ is a weight that equals the number of residents in state $s$  divided by the number of residents in HHS region $r$, and $\text{ILI}_{s}$ is the reported ILI for state $s$.
The influenza season begins on the epidemic week where the following three consecutive weeks are above a CDC defined baseline percentage of wILI~\cite{iliDef}. 
Details about the wILI dataset can be found in Supplemental~\ref{supp_ili}.

For computational convenience, ILI is discretized into 131 intervals: bins of the form $[X,Y)$ from 0\% to 13\% by 0.1\% and one bin of the form $[X,Y]$ from 13\% to 100\%.
We assume a sequence of random variables~(i.e. sample) $(X_{40}, X_{41}, \cdots, X_{20})$ are responsible for generating observed ILI values. 
No additional assumptions are made about this sample.
 
\subsection{Component model forecasts}
During the FluSight challenge, eight research teams (a subset of all teams) generated  twenty seven component model forecasts with the purpose of being combined into an ensemble called the FluSight Ensemble~\cite{reich2019collaborative}. 
Teams submitted weekly forecasts for seven targets associated with ILI and for seven influenza seasons that begin with the 2011/2012 season and end on the 2018/2019 season.
The seven targets are: one, two, three, four week-ahead percent ILI; season onset week (the week where three consecutive weeks’ ILI percentages are higher than the CDC baseline); season peak ILI percentage and season peak week. 

A component model forecast of 1-4 week ahead ILI at epidemic week $t$ is a discrete probability distribution over 131 intervals $[0,0.1], (0.1,0.2],(0.2,0.3],\cdots$ $(12.9, 13], [13,100]$. 
For `week' targets (season peak week and season onset week), each component model produces a probability distribution discretized over 32 epidemic weeks in a season.
In this work we only consider forecasts of 1-4 week ahead percent ILI.
See supplemental Fig.~\ref{supp.example_forecast} for an example of component model forecasts and Supp.~\ref{supp.componentModels} for a brief description of model types that were trained. 

\subsection{Comparator ensemble algorithms}
We compare our novel CAP-adaptive ensemble algorithm to three algorithms that have been implemented in past work: a equally weighted ensemble, a static ensemble, and an adaptive ensemble~\cite{mcandrew2021adaptively}.
All three ensembles are linear pools, assuming that observations are generated as a convex combination of component models.
\begin{align}
    \begin{aligned}
        f_{Y}(y | \pi_{1:C}) = \sum_{c=1}^{C} \pi_{c} f_{c}(y); \; 
        \forall \pi \geq 0; \; \sum_{c=1}^{C} \pi_{c}=1
    \end{aligned}
\end{align}
where $f_{c}$ is the predictive density over ILI values for component model $c$, $\pi_{c}$ is the weight assigned to component model $c$, there are $C$ component models considered, $y$ is percent ILI, and $f(y)$ is the predictive density of the ensemble. 
Because this is a multi-model ensemble, the component model densities $f_{c}$ are fixed. 
We have no access to parameters or model specifications for component models.

An equally weighted ensemble assigns component model weights to one over the number of component models.
A static ensemble assigns weights $\pi_{c}$ that maximize the log-likelihood 
\begin{align}
    \log \l[\ell( \pi_{1:C} | \mathcal{D}  ) \r] &= \sum_{j=1}^{N} \log \l[ f_{Y}(y_{j}) \r]  
                        = \sum_{j=1}^{N} \log \l[ \sum_{c=1}^{C} \pi_{c} f_{c}(y_{j}) \r] \label{LL_static}
\end{align}
where $\mathcal{D} = (y_{1},y_{2},\cdots, y_{N})$ is the sequence of observed ILI values from all previous seasons up to, but not including, the present season. 
During season $S$, weights are computed based on previous season ILI values and past component model forecasts.
Weights are assigned at the beginning of the season and stay fixed~(static) throughout the season.
Weights are then recomputed at the beginning of season $S+1$.
In the first season~(e.g. 2011/2012), with no past data, a static ensemble assigns equal weights to all component models. 
An adaptive ensemble~(ensemble 3) assigns equal weights to component models on week one and, week by week, updates weights. 
Weights~$(\pi_{1:C})$ are updated according to the log likelihood above~\eqref{LL_static} plus a time dependent Dirichlet prior.
The prior is meant to temper the model from assigning too much weight to a small number of component models~\cite{mcandrew2021adaptively}.

\section{CAP ensemble}

The CAP ensemble partitions (or clusters) the set of $C$ component models into a collection of $K$ sets of component models~(Cluster step). 
For cluster $k$,    map the set of component models to a single forecast~(Aggregate step) called a cluster forecast.
Then combine the $K$ cluster forecasts to a single CAP ensemble forecast~(Pool). 
The aim of the cluster and aggregate steps are to reduce component model redundancy.
The `Pool' step generates  a single forecast from a set of $K$ cluster forecasts.

\subsection{The impact of component model redundancy}
Consider a set of $C$ forecasts represented as random variables $X_{1},X_{2},\cdots,X_{C}$.
The linear pool forms a new random variable $Y$ with probability density function~(pdf) 
\begin{align}
    f(y) = \sum_{c=1}^{C} \pi_{c} f_{X_{c}}(y)
\end{align}
where $f_{X_{c}}(y)$ corresponds to the pdf for component model $c$.

\begin{figure}[ht!]
    \centering
    \includegraphics[width=1.\textwidth]{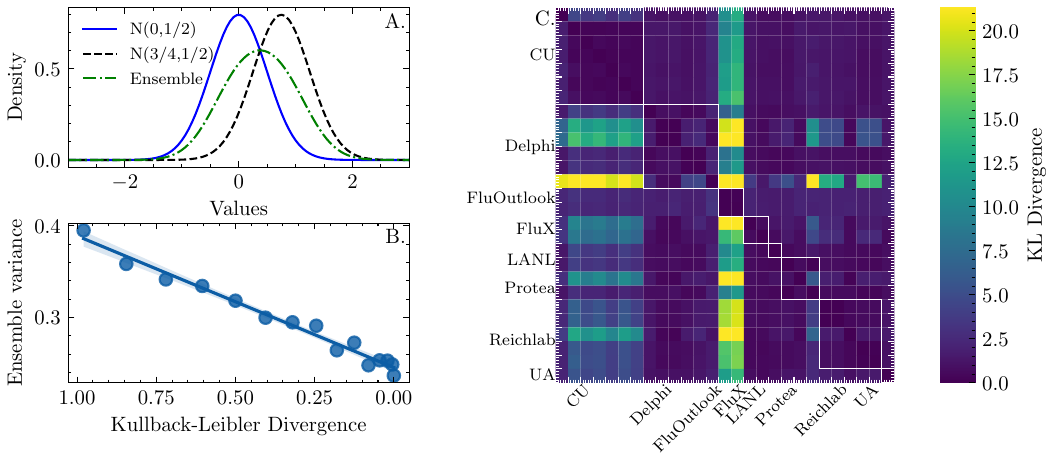}
    \caption{(A.)~An example of an equally-weighted ensemble of two non-overlapping component models. The variance of the ensemble is greater than the component models because they have different expected values.
    (B.)~Kullback Leibler divergence between two component models~(one model produced a Normal density with expected value 3/4 and variance 1. The second model has a variance of 1 and the expected value moves sequentially from, 3/4 to 3/2 by 0.05) vs the variance of an equally weighted ensemble of the two models.
    (C.)~Pairwise KL divergence for all 27 component models averaged over 11 locations and 4 `week ahead' targets. Models corresponding to teams are labeled with text and a white box. 
    Overlap (i.e. redundancy) artificially reduces the variance of an ensemble forecast, producing a forecast that is too sharp.
    An ensemble where component models have little overlap should produce a forecast that is not too narrow.}
  \label{fig.var_reduction}
\end{figure}

Similar, or overlapping, densities submitted by more than one component models may present issues with: (i) forecast variance and (ii) identifiability. 
We can illustrate these two issues by combining overlapping component models with the following example. 
Suppose we consider combining in an equally weighted linear pool two component models represented by the below two random variables
\begin{align}
    X_{1} \sim \mathcal{N}\left(\mu_{0},\sigma^{2} \right);\;
    X_{2} \sim \mathcal{N}\left(\mu_{1},\sigma^{2} \right);\;
\end{align}
Then the variance of the linear pool ensemble~(see \cite{hogg2019introduction} for a detailed derivation) will be 
\begin{align}
   \mathbb{V} &= \sum_{c=1}^{2} \frac{1}{2} \sigma_{c}^{2} + \sum_{c=1}^{2} \frac{1}{2} \mu_{c}^{2} - \left[\sum_{c=1}^{2} \frac{1}{2}  \mu_{c} \right]^{2}
\end{align}
Because squaring is a convex function, we can appeal to Jensen's inequality to find that  
\begin{align}
    \sum_{c=1}^{2} \frac{1}{2} \mu_{c}^{2} > \left[\sum_{c=1}^{2} \frac{1}{2}  \mu_{c} \right]^{2}
\end{align}
If the expected values are equal than the variance reduces to $\sigma^{2}$, but if the expected values are distinct than the linear pool variance will be greater than $\sigma^{2}$~(Fig~\ref{fig.var_reduction}A.).
We find that as the Kullback-Leibler~(KL) divergence between the two component model probability density functions decreases~(i.e. the two cmponent model forecasts become more similar) the ensemble variance of an equally weighted ensemble decreases linearly~(Fig~\ref{fig.var_reduction}B.).  
The 27 component models submitted as part of the FluSight Ensemble have on average a low KL divergence which indicates a high level of redundant/similar model forecasts~(Fig~\ref{fig.var_reduction}C.). 

\begin{figure}[ht!]
    \centering   \includegraphics{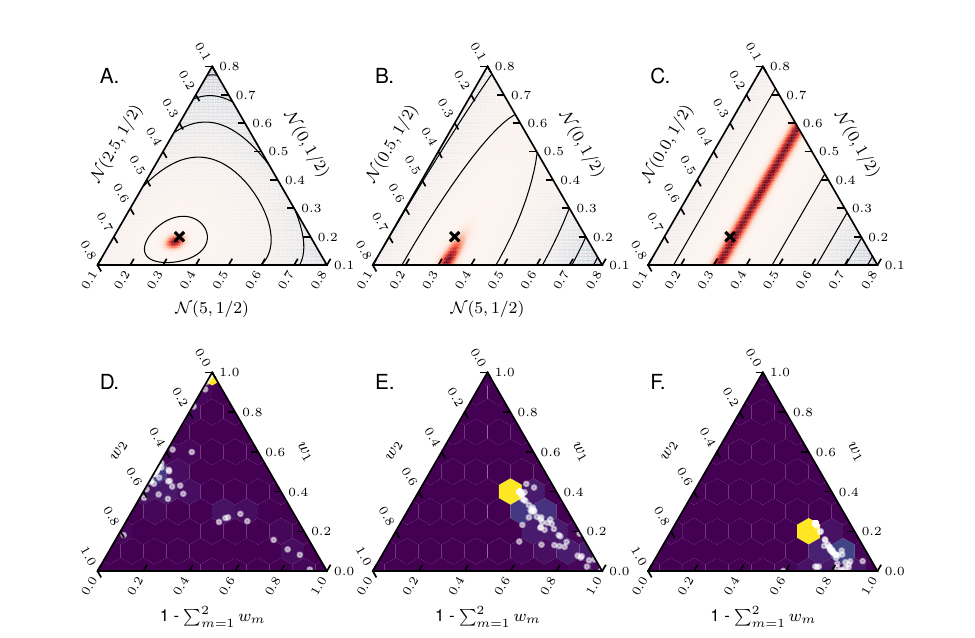}
    \caption{(A.)-(C.)~ Loglikelhioood surface~(contour lines) for a linear pool that combines three component models under three levels of redundancy. The black $X$ marks the true weights.
    Red shading indicates a multivariate approximation to the estimated optimal weight assignment and variability around that estimate. 
    (A.)~ Little redundancy between component models, (B.) moderate redundancy, and (C). two component models that are identical.
    (D.)-(F.)~
    From the FluSight challenge we chose three empirical examples~(D.~Season 2016/2017, Week 10, HHS Region 9, 1 week ahead forecast; E.~Season 2014/2015, Week 16, HHS 8, 4 week ahead forecast; F.~Season 2014/2015, Week 46, US, 1 week ahead forecast) of optimal weight assignments~(white circles) using 100 random restarts for two component models~($w_{1}$ and $w_{2}$) and weights for the remaining 25 component models from FluSight.
    \label{fig.t_and_r}}
\end{figure}

Redundancy also presents an issue if the modeler decides to estimate weights for an ensemble~(Fig.~\ref{fig.t_and_r}). 
If there exists two models with a large overlap in predictive densities then the log-likelihood surface is flat~(i.e. the determinant of the Hessian at the global optimum is near zero) near the global optimum~(Fig.~\ref{fig.t_and_r}A.-C.). 
When training ensemble weights with a standard optimizer, random restarts will show that the optimal weight vector has high variance. 
High variance in an optimal solution is characteristic of non-indentifiability.
Non-indentifiability issues are present in FluSight component models~(Fig.~\ref{fig.t_and_r}D.-F.).

Our proposed CAP adaptive algorithm identifies component models whose predictions produce similar scores (logscores) and groups them into one cluster for further aggregations. This approach aim to enhance the diversity of predictions within the ensemble and potentially leadsto more robust and accurate overall forecating results.

\clearpage
\subsection{CAP framework}

Below we present the Clustering, Aggregation, and Pooling steps for the CAP ensemble and specific choices we made to compare a CAP to non-CAP ensemble.

\subsubsection{Clustering step}
The goal of the clustering step is to partition $C$ component model forecasts $f_{1},f_{2},\cdots, f_{C}$ into a collection $\Gamma = \{ c_{1}, c_{2}, \cdots, c_{K} \}$ where the $i^{\text{th}}$ cluster, $c_{i}$, is defined as a set of component models forecasts.
Each component model forecast may belong to only one cluster.
There are no other constraints to implementing the clustering step, and want to stress that the reader/user should be free to choose their own algorithm for this step.
Clustering algorithms may arise from a case-by-case intuition, professional expertise, or data-driven results.
A specific example of such an algorithm is below.

We chose a heuristic method to cluster component models. 
Component model $m$ is placed in cluster $c$ if this model has a linear correlation coefficient above a threshold value $\phi$ for all models in cluster $c$.
If model $m$ does not satisfy the above criteria for any current clusters then we generate a new cluster with the single model $m$ and continue the algorithm.
We use a data-driven approach to choose $\phi$.
Let $t=1$ equal the first epidemic week in a season, $t=2$ the second week until $t=T$ where $T$ is the last epidemic week in the season. 
If $t=1$ set a threshold value $\phi$ equal to $1/2$.
If $t>1$: then, given a value $\phi$, complete the Aggregate/Pool~(see below for the options we chose) steps to generate ensemble forecasts over all regions and targets for which there exists ground truth percent ILI. Average the logscores for this choice of $\phi$. 
Select the value $\phi$ with the highest average logscore.

\subsubsection{Aggregating step}
The goal of the aggregation step is to produce $K$ forecasts---called cluster forecasts---by combining the component models that belong to each cluster into a single predictive density over future values of ILI. 
The aggregation step is a choice of function $h$ that for each cluster $c_{k}$, maps component model predictive densities $(f_{(c_{k},1)},f_{(c_{k},2)},\cdots,f_{(c_{k},s_{k})})$ belonging to a cluster into a single predictive density $F_{c_{k}}$ 
\begin{align}
    F_{c_{k}}  = h(f_{(c_{k},1)},f_{(c_{k},2)},\cdots,f_{(c_{k},s_{k})})
\end{align}
where $s_{k}$ equals the number of component models that belong to cluster $c_{k}$.
Again, we can stress that the reader/user is free to choose any aggregation algorithm.

For our example CAP ensemble, the aggregation algorithm we proposed---sometimes called a `follow the leader' approach---computes the median log score assigned to all past forecasts for each component model that belongs to one cluster and then selects as the cluster forecast the component model with the highest median log score.

\subsubsection{Pooling Step}
The goal of the pooling step is to assign to $K$ cluster forecast predictive densities a vector of weights $\pi = [\pi_{1}, \pi_{2}, \cdots, \pi_{K}]'$ that optimizes an objective function. 
We chose two approaches: (i)~equal weights or (ii)~an adaptive weighting scheme that is identical to the adaptive ensemble approach but applied to cluster, instead of component model, forecasts.

\section{Forecast Evaluation}

Ensemble forecasts were evaluated by using the logarithmic score, probability integral transform~(PIT) value, and the Brier score.

The logarithmic score~(log score) was used to evaluate forecasts~\cite{winkler1968good,gneiting2007strictly}.
The logarithmic score is defined as the natural log of the predicted probability assigned to the eventually realized true ILI value.

Given a model with probability density function
$f(z)$ and a true ILI value $i*$, the log score was defined as log of the probability assigned to the discretized bin $[a,b)$ which contains the true wILI value
\begin{equation}
    \text{log score}(f)=\int_{a}^{a+0.1\%} f(z)dz,
\end{equation}
where $0.1\%$ is the bin width. 
The log score approaches negative infinity as the probability a forecast assigns to the truth approaches zero, and the maximum~(best) log score is a value of zero.
Logscores that were smaller than -10 were set to -10 to replace any extreme small log scores, as was common practice during the ILI FluSight challenges~\cite{mcgowan2019collaborative,reich2019collaborative}.

The probability integral transform~(PIT) value is defined as the probability that a forecaster assigns to values that are less than or equal to the ground truth $t$. 
Suppose a forecaster submits a predictive density $f$. Then the PIT value is computed as 
\begin{align}
    \text{PIT}(f,t) = \int_{-\infty}^{t} f(x) \; dx.
\end{align}
If a forecaster is perfectly calibrated, then out of 100 ground truth values we would expect that they assign the PIT value of $x$ to $x \times 100$ of these ground truth values~\cite{gneiting2007strictly}.

The PIT value can identify if a forecaster is too confident about the future, too unsure, too often assigns values larger than the truth or vice versa~\cite{gneiting2007strictly}.
If a forecaster is too sure about the future then they will submit a narrow predictive density, which will cause PIT values to be frequently small and close to zero or large and close to one.
If a forecaster is too unsure about the future, then they will submit a wide predictive density causing PIT values to be frequently close to 0.50. 
If a forecaster frequently over estimates the truth, then they will submit a density with a median above the truth, which will cause PIT values to be frequently smaller than 0.50 and few PIT values to be larger than 0.50. 
If a forecaster frequently underestimates the truth, then we expect that the median will be smaller than the truth, which will cause PIT values to often be larger than 0.50.

The Brier score is defined as 
\begin{align}
    \text{BrS}(F,t,x) = \l[ F(x) - \mathbbm{1}( x < t ) \r]^{2}
\end{align}
where $F$ is the predictive density, $t$ the true percent ILI, and $x$ a threshold value~\cite{brier1950verification}.
We computed the Brier score over threshold values from 0 percent ILI to 10 percent ILI by steps of size 0.1 percent.
To note, integrating the Brier score over all possible threshold values is equal to the continuous rank probability score~(or CRPS)~\cite{gneiting2007probabilistic}.

\section{Results}

\subsection{Component model performance and correlation over an influenza season}
\begin{figure}[ht!]
    \centering
    \includegraphics[width=1.\textwidth]{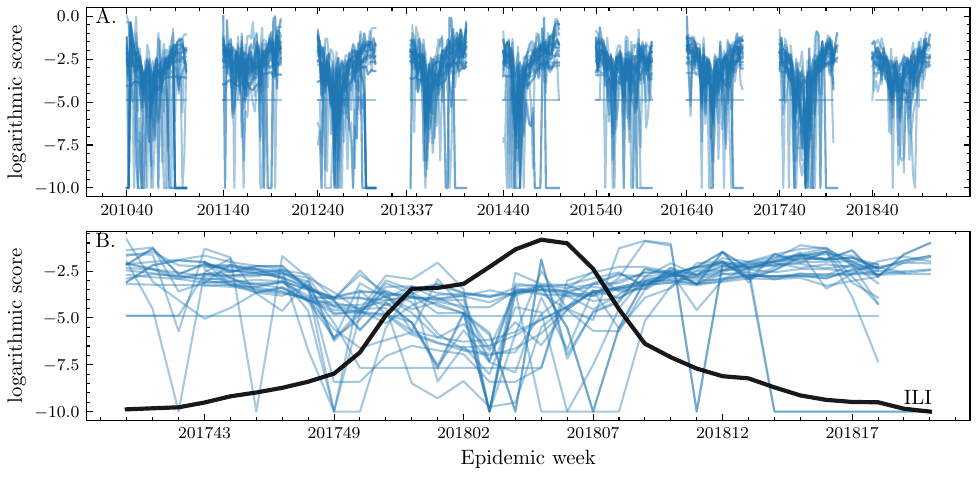}
    \caption{ (A.)~Logarithmic scores~(log scores) for two-week ahead forecasts~(`nowcasts') of percent ILI at the US national level produced by all 27 component models from the 2010/2011 season until the 2018/2019 season. 
    (B.)~Logscores for US national nowcasts generated by component models  in the 2017/2018 season, and in black is the US national estimate of percent ILI.
    Component model log scores show a characteristic pattern throughout a season, suggesting that component model structures may be redundant. \label{fig.logscore}}
\end{figure}

Component model performance as measured by log score oscillates from the beginning of the 2010/2011 season until the end of the 2018/2019 season~(Fig.~\ref{fig.logscore}A.). 
Component model log scores are highest~(best) during the beginning and end of a season and perform worst during the middle of the season, close to peak ILI~(Fig.~\ref{fig.logscore}B.).
The observation that log scores similarly decrease during a peak suggests that many component models may provide similar, redundant forecasts~(See supplemental Fig.~\ref{fig.corrAndili} of pairwise correlations between component models at four points during the 2017/2018 season).

\subsection{Clustering component models by log scores}

\begin{figure}[ht!]
    \centering
    \includegraphics[width=1.0\textwidth]{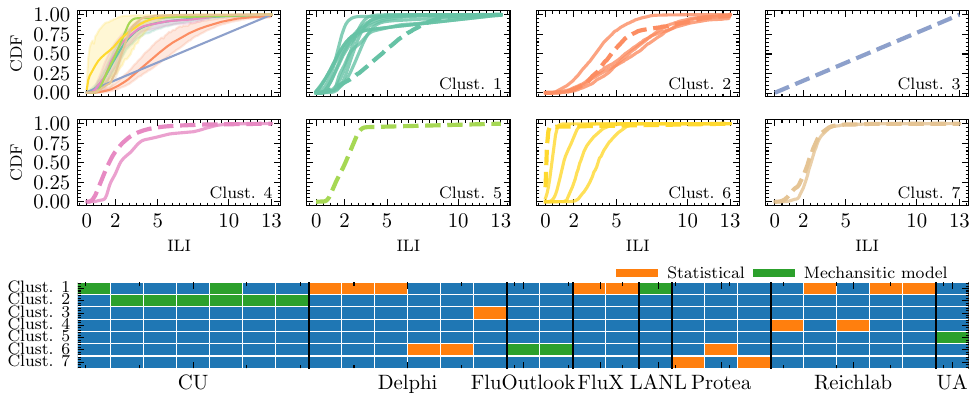}
    \caption{(Top.)~Component model cumulative density functions submitted as 3-week-ahead forecasts for HHS region 1, MW114, grouped into seven clusters based on the log score. 
    The top left panel computes for each cluster the mean CDF value and 95\%CI. 
    CDF lines that are dashed correspond to those models that were chosen~(based on highest median log score within cluster) to be combined into a final CAP ensemble forecast. 
    (Bottom)~ Cells that are filled with green/orange 
    Rows correspond to clusters and columns correspond to component models. 
    Black vertical lines group together component models by research group.
    If a cell is filled with orange/green in row $r$ and column $c$  then component model $c$ belongs to cluster $r$.
    Orange cells indicate a component model that is statistical and green indicates a mechanistic model.~\label{example_cluster}}
\end{figure}

The CAP ensemble uses past log scores to group together component models with similar predictive densities over 1, 2 , 3, and 4 week ahead future ILI~(Fig.~\ref{example_cluster}).
The choice to aggregate a cluster by choosing the component model with the highest median logscore appears, for most clusters, representative of the other component model forecasts belonging to the that cluster~(dashed line in Fig.~\ref{example_cluster}).
An extra advantage of clustering by logscore is that component models within one cluster appear to have similar structural properties~(Fig.~\ref{example_cluster} bottom).
%

\subsection{Comparison of performance between CAP and Non-CAP ensembles}

\begin{figure}[ht!]
    \centering
    \includegraphics[width=\textwidth]{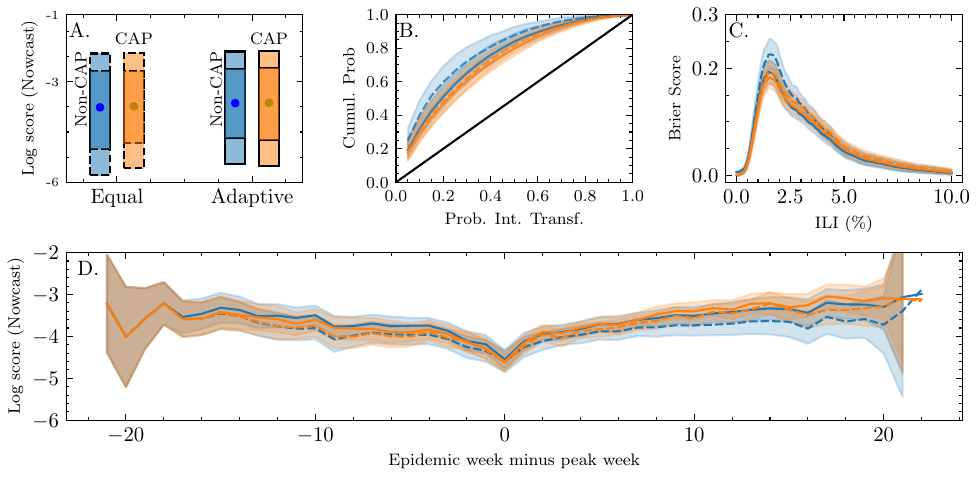}
    \caption{ The CAP algorithm~(orange) outperforms a non-CAP~(blue) for an equally weighted~(dashed line) and adaptive~(solid line) weighted ensemble.
    Performance metrics are aggregated across region and for the two-week ahead "Nowcast" prediction.
    (A.)~The median, 10 and 90, 25 and 75th percentiles of logscores.
    (B.)~Cumulative density function of probability integral transform~(PIT) values.
    (C.)~Brier scores computed by partitioning forecasts above and below a sequence of ILI "cutpoints" from 0 to 10.
    (D.)~Logscore is presented on the vertical axis and the number of weeks before and after the peak ILI.
    \label{fig.all_scores}}
\end{figure}

The CAP algorithm---compared to a non-CAP approach---tends to improve calibration and has similar accuracy~(Fig.~\ref{fig.all_scores}).
Compared to a non-CAP algorithm, applying the CAP algorithm to an equally weighted and adaptive ensemble shows a similar distribution of logscores (Fig.~\ref{fig.all_scores}A.).
The distribution of PIT values and Brier scores show that the CAP algorithm produced less biased~(i.e. overestimates ILI less) forecasts compared to non-CAP~(Fig.~\ref{fig.all_scores}B. and C.).
The average area under the curve between the CDF for PIT values and the identify line~(smaller values are better) was, for an equally weighted ensemble $0.25$ for a CAP and 0.30 non-CAP~($\sim$ 17\% improvement) and for an adaptive weighted ensemble $0.24$ for CAP and 0.26 non-CAP~($\sim$ 8\% improvement).
The average integral over Brier scores from an ILI value of zero to 10 percent~(smaller is better) is, for an equally weighted ensemble $0.66$ for CAP and $0.69$ non-CAP, and for an adaptive weighted ensemble $0.61$ for CAP, $0.60$ non-CAP. 
Applying a CAP tends to have similar accuracy at peak ILI, worse accuracy at the beginning of the season (i.e. at the beginning of the flu season), and improved accuracy after the peak~(Fig.~\ref{fig.all_scores}D.).
The CAP algorithm---compared to a non-CAP and aggregating over equal or adaptive weighting---improves calibration~(i.e. PIT) by $\sim$10\% and has a close to zero difference in logscore and Brier score. 

%

\clearpage
\section{Number of clusters and distribution of weights associated with CAP algorithm}

\begin{figure}[ht!]
    \centering
    \includegraphics{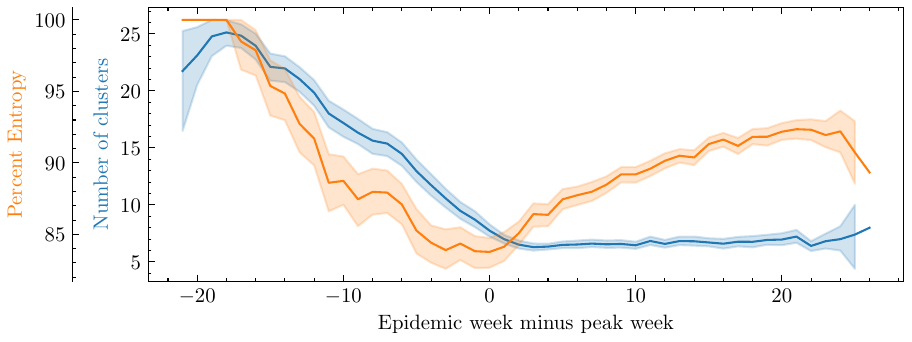}
    \caption{The number of clusters and the percent entropy (defined as entropy of weights assigned to clusters divided by maximum entropy) for the CAP-Adaptive ensemble aggregated over all locations and week-ahead targets.
    The large number of clusters before the peak and small number after the peak reflects the difficulty in forecasting before peak ILI. 
    The decrease in percent entropy before and at the peak suggests that the CAP algorithm presented is learning which models perform best. However, the increase in percent entropy after the peak could be because of varied performance during peak ILI. 
    \label{fig.alg_extras}}
\end{figure}

For the proposed CAP-adaptive ensemble algorithm, the number of clusters is on average 23 at the beginning of the influenza season~(20 weeks or more before peak), 8 during peak ILI~(within one week of peak), and 7 by the end of the season~(20 weeks or more after the peak)~(Fig.~\ref{fig.alg_extras} blue).
The percent entropy is on average 100\% at the beginning of the season, 83\% at the peak, and 92\% by the end of the season(Fig.~\ref{fig.alg_extras} orange).

\section{Discussion}
A Combine-Aggregate-Pooling~(CAP) approach has the potential to improve forecatsing performance of 1-4 week ahead forecasts of ILI compared to an equally weighted, static, and adaptive ensemble approach. 
In particular, a CAP approach was observed to be better calibrated when compared to a non-CAP approach. 

We feel that a major advantage to a CAP approach is the potential to reduce redundancy in component models. 
Redundant component models can cause issues of non-identifiability 
which may produce a smaller~(than expected) linear pool variance and make it difficult, or impossible, to learn component model weights.
By clustering component models and selecting from each cluster a single representative component models, the CAP approach is able to reduce redundancy between component models.

A `built-in' advantage of applying a CAP approach is the ability to handle missing forecasts. 
In real-time forecasting efforts, a schedule is typically assigned for when to submit component models forecasts.
Forecasts may not be submitted successfully because a change in the data breaks a component model algorithm, the modeling team misses the deadline due to other priorities, the generated forecast does not match intuition, etc.
Past ensemble approaches accounted for missing forecasts in an adhoc fashion, using methods that were separate from the ensemble approach~\cite{reich2019accuracy,mcandrew2021adaptively}.
Because a CAP approach combines cluster forecasts and not component model forecasts there is a small probability that a cluster forecast is missing.
A cluster forecast is missing only if every every component forecast that is included in the cluster is missing. 

Another advantage to CAP is the ability to more easily include public health officials in the ensemble procedure. 
For past ensemble models, public health officials could participate by assigning apriori weights to component models. 
However, assigning weights may not be as intuitive as grouping together component models.
The CAP approach allows public health officials~(PHOs) to engage in ensemble building by grouping component models if there exists an important reason to do so. 
Clustering does not need to be performance based but could be instead based upon how the PHO wants to communicate information.
For example, the PHO may decide that all models associated with a research group will be clustered together or all models of a specific type~(statistical, mechanistic, etc) may be grouped together. 
Forecasts from each cluster and an ensemble forecast can be presented to the PHO group who then disseminates this information to the public.

In this work we chose a specific implementation of CAP, however the CAP approach is broad and can include many methods for combining component models, aggregating models, and pooling.
In future work we wish to explore several implementations of the CAP to determine if there exists an optimal combination of these three steps.  
We also wish to explore training a CAP ensemble on forecasts that are formatted as a sequence of quantiles instead of a discretized probability mass function.
The quantile format has become the more popular option to format forecasts and the CAP ensemble approach should be expanded to combine forecasts in this format.

Our specific CAP implementation and the above results present limitations. 
The combine and aggregate steps for our ensemble use historical logscores for component models. 
Here, we assume that a high correlation between log scores of two component models implies that these models must share similar modeling structure and predictive density functions. 
This may not always be the case.
In addition, readers may expect that clustering should partition component models based on structural attributes~(such as phenomenological vs mechanistic, team producing the model, etc) but this is not always the case~(Fig.~\ref{example_cluster}).
Only two weighting procedures were evaluated: an equal and adaptive weighting procedure. 
Other weighting schemes should be evaluated.
Another limitation is that this procedure was evaluated on the older discretized probability format for forecasts.
The CAP procedure still, as a last step, pools cluster forecasts, and so assumes that cluster forecast performance should be consistent over the season. This is a limitation because cluster performance may not be consistent over time just as component model performance can be inconsistent.

\section{Acknowledgements}
We wish to thank the CMU Delphi group for providing easy, programmatic access to ILI data~(\url{https://cmu-delphi.github.io/delphi-epidata/
}). We wish to thank Nicholas G. Reich for running the FluSight Network project which allowed for a rich dataset of forecasts~(\url{https://github.com/FluSightNetwork/cdc-flusight-ensemble}).

\section{Data Availability}
ILI data used for this manuscript can be located at \url{https://cmu-delphi.github.io/delphi-epidata/api/flusurv.html}. 
A script titled ``download\_epidata.py'' at \url{https://github.com/computationalUncertaintyLab/CombineAggregatePool/} can be used to download all ILI data.
Forecast output for all component models can be found at \url{https://github.com/FluSightNetwork/cdc-flusight-ensemble/tree/master/model-forecasts/component-models} and a script titled `combineFSNForecastsTogether.py' at \url{https://github.com/computationalUncertaintyLab/CombineAggregatePool/} can be used to combine all forecasts into a single dataset.
Algorithm and figures code can be found at \url{https://github.com/computationalUncertaintyLab/CombineAggregatePool/}.




\section{Supplemental}

\section{Influenza-like illness by season}

\begin{figure}[ht!]
    \centering
    \includegraphics[width=\textwidth]{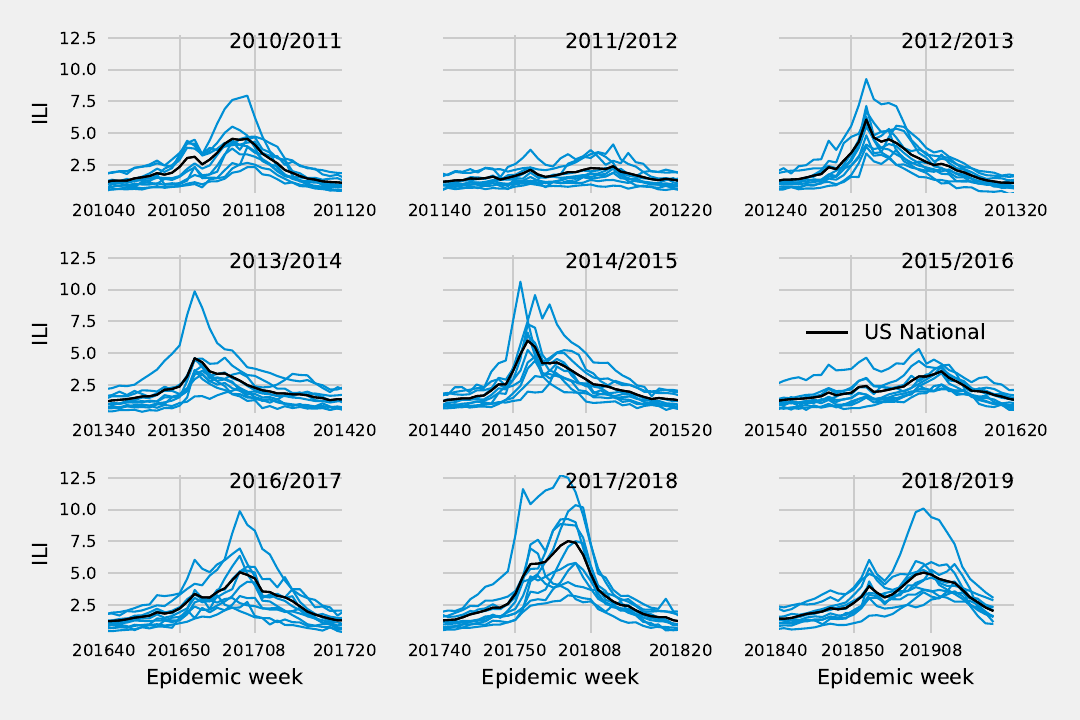}
    \caption{Percent influenza-like illness~(ILI) values for HHS regions 1-10~(blue) and a US national estimate~(black) from the 2010/2011 season to the 2018/2019 season.
    ILI values across regions follow a similar pattern within one season. Across seasons, ILI follows different trajectories.   
    \label{supp.ili}}
\end{figure}

\section{Component model forecasts}

We organize the component forecasts of 27 component model for all HHS regions, all forecast targets, and all epidemic weeks as
\begin{align}
    \mathcal{F} = \begin{bmatrix}
      \text{Region} & \text{Target} & \text{Component Model ID} & \text{EW} & \text{bin 1} & \text{...} & \text{bin 131}  \\ 
      \text{HHS1} & \text{1} & \text{1} & 201040 & \text{XXX} & ... & \text{XXX}\\
      \vdots\\
      \text{HHS10} & \text{4} & \text{27} & 201820 & \text{XXX} & ... & \text{XXX}\\
      \text{Nat} & \text{1} & \text{1} & 201040 & \text{XXX} & ... & \text{XXX}\\
      \vdots & \vdots & \vdots & \vdots & \vdots & \vdots & \vdots\\
       \text{Nat} & \text{4} & \text{27} & 201920 & \text{XXX} & ... & \text{XXX}\\
    \end{bmatrix}
\end{align}
Each row of $\mathcal{F}$ is a discretized forecast distribution of a component model with specified target and region. 131 ILI bins are ordered in ascending order. Each contains a probability of the forecasted ILI value made by the component model. For example, at epidemic week 201040, component model id 0 (CUBMA) generates a discretized distribution (row 1, bin 1 $\sim$ bin 131) of predicted wILI value for epidemic week 201041 in region HHS1.

\begin{figure}[ht!]
    \centering
    \includegraphics[width=1\textwidth]{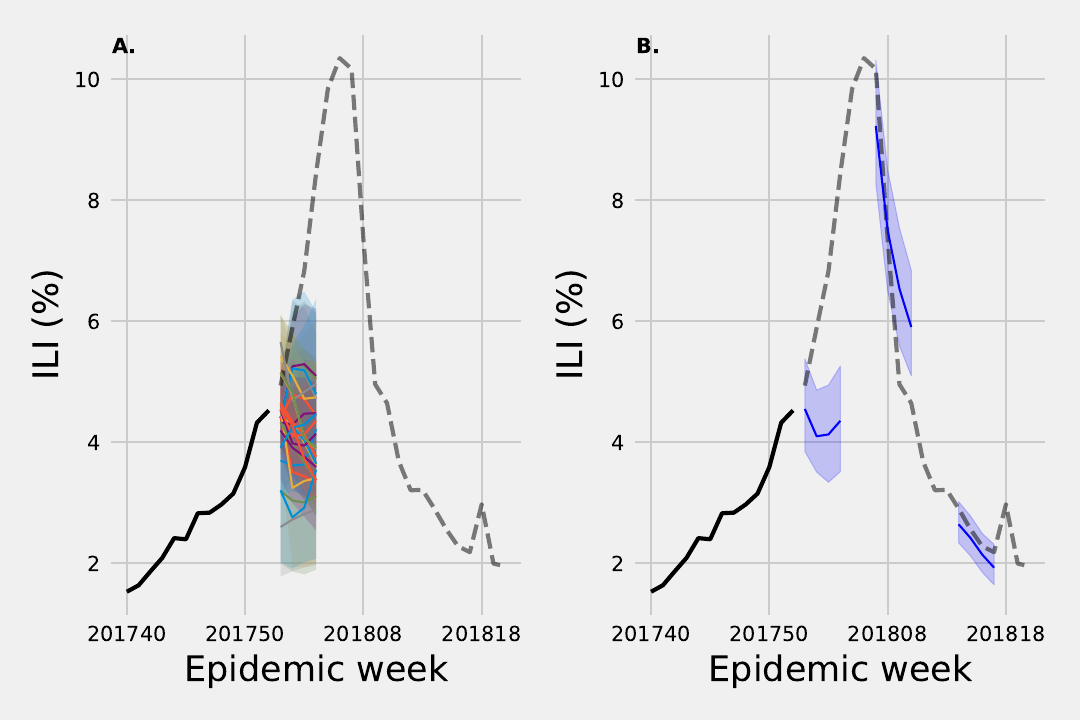}
    \caption{(A.)~Component model forecasts one through four weeks ahead at epidemic week 51 in HHS 3 for the 2017/2018 season represented as a median (solid line) plus 25th, 75th quantiles~(shaded region). (B.)~A single component model forecast one through through four weeks ahead at three different epidemic weeks.
    The solid line represents ground truth ILI data that is observed by all models. 
    The dotted line represents ground truth ILI that models have not yet observed.
    Component model forecasts are open source and available at~\cite{FSN} in the folder ./model-forecasts/component-models/. 
  \label{supp.example_forecast}}
\end{figure}

\clearpage
\section{Correlation in log score over time}
\begin{figure}[ht!]
    \centering
    \includegraphics[width=\textwidth]{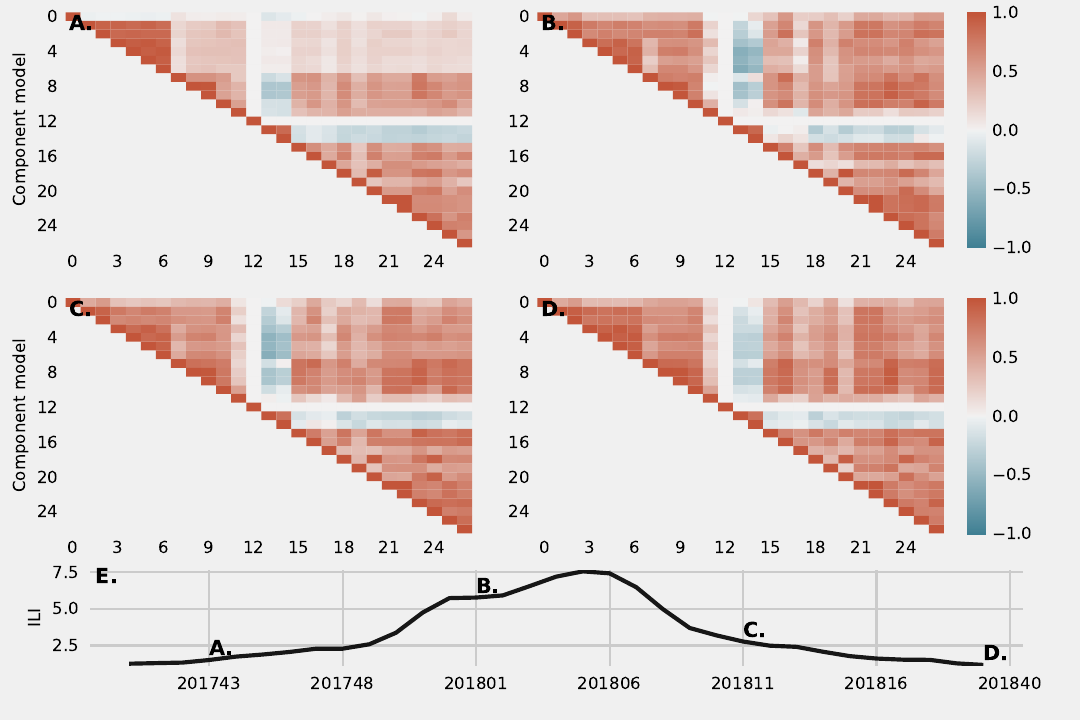}
    \caption{Pairwise linear correlation for 2 week ahead forecasts of ILI among all 27 component models at four different epidemic weeks:(A.)~201743 (B.)~201801 (C.)~201811 (D.)~201820. (E.) ILI for the 17/18 season.~\label{fig.corrAndili} }
\end{figure}

\section{Descriptions of twenty-seven Component models}

\begin{table}[ht!]
\begin{tabular}{||c|c|c|c||} 
 \hline
 Model ID & Component Model & Research Team & Submission Type\\ [0.5ex] 
 \hline\hline
 1 & BMA & CU &real-time \\ 
 \hline
 2 & EAKFC-SEIRS & CU &real-time \\ 
 \hline
 3 & EAKFC-SIRS & CU &real-time \\
 \hline
 4 & EKF-SEIRS & CU &real-time \\
 \hline
 5 & EKF-SIRS & CU &real-time \\
 \hline
 6 & RHF-SEIRS & CU &real-time \\
 \hline
 7 & RHF-SIRS & CU &real-time \\
 \hline
 8 & Basis Regression & Delphi & real-time\\
 \hline
 9 & Empirical Futures & Delphi & real-time\\
 \hline
 10 & Empirical Trajectories & Delphi & real-time\\
 \hline
 11 & Delta Density & Delphi & real-time\\
 \hline
 12 & Markovian Delta Density & Delphi & real-time\\
 \hline
 13 & Uniform Distribution & Delphi & real-time\\
 \hline
 14 & Mechanistic GLEAM Ensemble & FluOutlook & retrospective\\
 \hline
 15 & Augmented Mechanistic GLEAM Ensemble & FluOutlook & retrospective\\
 \hline
 16 & Auto Regressive model with Likelihood Ratio based Model Selection & FluX & retrospective\\
 \hline
 17 & Long Short-Term Memory~(LSTM) based deep learning & FluX & retrospective\\
 \hline
 18 & Dynamic Bayesian Model plus & LANL & real-time\\
 \hline
 19 & Ensemble of dynamic harmonic model and historical averages & Protea & real-time\\
 \hline
 20 & Subtype weighted historical average model & Protea & real-time\\
 \hline
 21 & Dynamic Harmonic Model with ARIMA errors & Protea & real-time\\
 \hline
 22 & Kernel Conditional Density Estimation & ReichLab & real-time\\
 \hline
 23 & Kernel Density Estimation & ReichLab & real-time\\
 \hline
 24 & Kernel Conditional Density Estimation with post-hoc backfill adjustment & ReichLab & retrospective\\
 \hline
 25 & SARIMA model without seasonal differencing & ReichLab & real-time\\
 \hline
 26 & SARIMA model with seasonal differencing & ReichLab & real-time\\
 \hline
 27 & Epidemic Cosine with Variational Data Assimilation & UA & retrospective\\
 \hline
\end{tabular}
    \caption{Brief description of twenty-seven component models}
    \label{supp.componentModels}
  \end{table}

\end{document}